\documentclass{Interspeech}

\interspeechcameraready

\title{Mitigating Subgroup Disparities in Multi-Label Speech Emotion Recognition: A Pseudo-Labeling and Unsupervised Learning Approach}

\author[affiliation={1}]{Yi-Cheng}{Lin}
\author[affiliation={2}]{Huang-Cheng}{Chou}
\author[affiliation={1}]{Hung-yi}{Lee}

\affiliation{Graduate Institute of Communication Engineering}{National Taiwan University}{Taiwan}
\affiliation{}{Independent Researcher}{Taiwan}
\email{\{f12942075, hungyilee\}@ntu.edu.tw, huangchengchou@gmail.com}

\usepackage{comment}
\usepackage{bbm}
\usepackage{hyperref}
\usepackage{amsmath}
\usepackage{graphicx}
\usepackage{subcaption}
\usepackage{booktabs}
\usepackage{multirow}
\usepackage{tikz}
\usepackage{wasysym}
\usepackage{pgfplots}
\pgfplotsset{compat=1.17} 
\usepackage{amsmath}
\usepackage{times} 
\usepackage{colortbl}

\begin{document}

\maketitle

\begin{abstract}
    
    While subgroup disparities and performance bias are increasingly studied in computational research, fairness in categorical \emph{Speech Emotion Recognition} (SER) remains underexplored. Existing methods often rely on explicit demographic labels, which are difficult to obtain due to privacy concerns. To address this limitation, we introduce an Implicit Demography Inference (IDI) module that leverages pseudo-labeling from a pre-trained model and unsupervised learning using k-means clustering to mitigate bias in SER. Our experiments show that pseudo-labeling IDI reduces subgroup disparities, improving fairness metrics by over 28\% with less than a 2\% decrease in SER accuracy. Also, the unsupervised IDI yields more than a 4.6\% improvement in fairness metrics with a drop of less than 3.6\% in SER performance. Further analyses reveal that the unsupervised IDI consistently mitigates race and age disparities, demonstrating its potential when explicit demographic information is unavailable.

\end{abstract}


\section{Introduction}


\emph{Speech Emotion Recognition} (SER) is essential in human–computer interaction, enhancing user experiences across various applications \cite{Thiripurasundari_2024}. As SER becomes increasingly integral to applications ranging from virtual assistants to therapeutic tools, ensuring that these systems operate fairly and without bias is more critical than ever \cite{Daniel_2022, Carolyn_2023}. However, conventional debiasing methods predominantly depend on explicit demographic annotations (e.g., gender, age, race), which are not only challenging to collect due to privacy concerns and ethical issues but are also notably scarce in real-world data \cite{Lin_2024, Gorrostieta_2019, lin24b_interspeech, Chien_2023, Chien_2024, 10446326, zhang22n_interspeech}.


Existing gender debiasing methods in SER are scarce, particularly those that do not require gender labels. Additionally, prior approaches have been designed for single-label categorical SER, while recent research highlights the importance of multi-label SER \cite{Chou_2024_v3, Wu_2024}, aligning with psychological studies on emotion co-occurrence \cite{Cowen_2021}. Against these backdrops, our research question is: \textbf{\emph{How can we mitigate subgroup disparities in multi-label SER without relying on explicit demographic annotations}?}


To address this challenge, we introduce the concept of \emph{Implicit Demography Inference (IDI)}. This novel framework aims to simulate demographic supervision by inferring latent group cues directly from the speech data. The IDI module operates through two techniques: first, by leveraging pseudo-labeling to generate proxy demographic labels using a pre-trained model, and second, by employing unsupervised clustering—specifically, applying K-Means to ECAPA-TDNN embeddings \cite{Desplanques_2020}—to uncover inherent group structures. Together, these strategies enable the system to approximate demographic distinctions without requiring manual annotations.

Our experiments on the CREMA-D database \cite{Cao_2014} provide compelling evidence for the efficacy of this approach. We observe that pseudo-labeling yields improvements in key fairness metrics by over 28\%, while the unsupervised clustering approach enhances these metrics by approximately 4.6\%, all with only a minor compromise in overall SER accuracy. Furthermore, the SER model debiased using the unsupervised clustering method can also suppress age and race bias, proving the generalizability of our method. These results underscore the potential of the IDI module to serve as a robust and scalable solution for debiasing in SER, especially in environments where explicit demographic data is limited.

\begin{figure}[t]  
    \centering
    \includegraphics[width=0.45\textwidth]{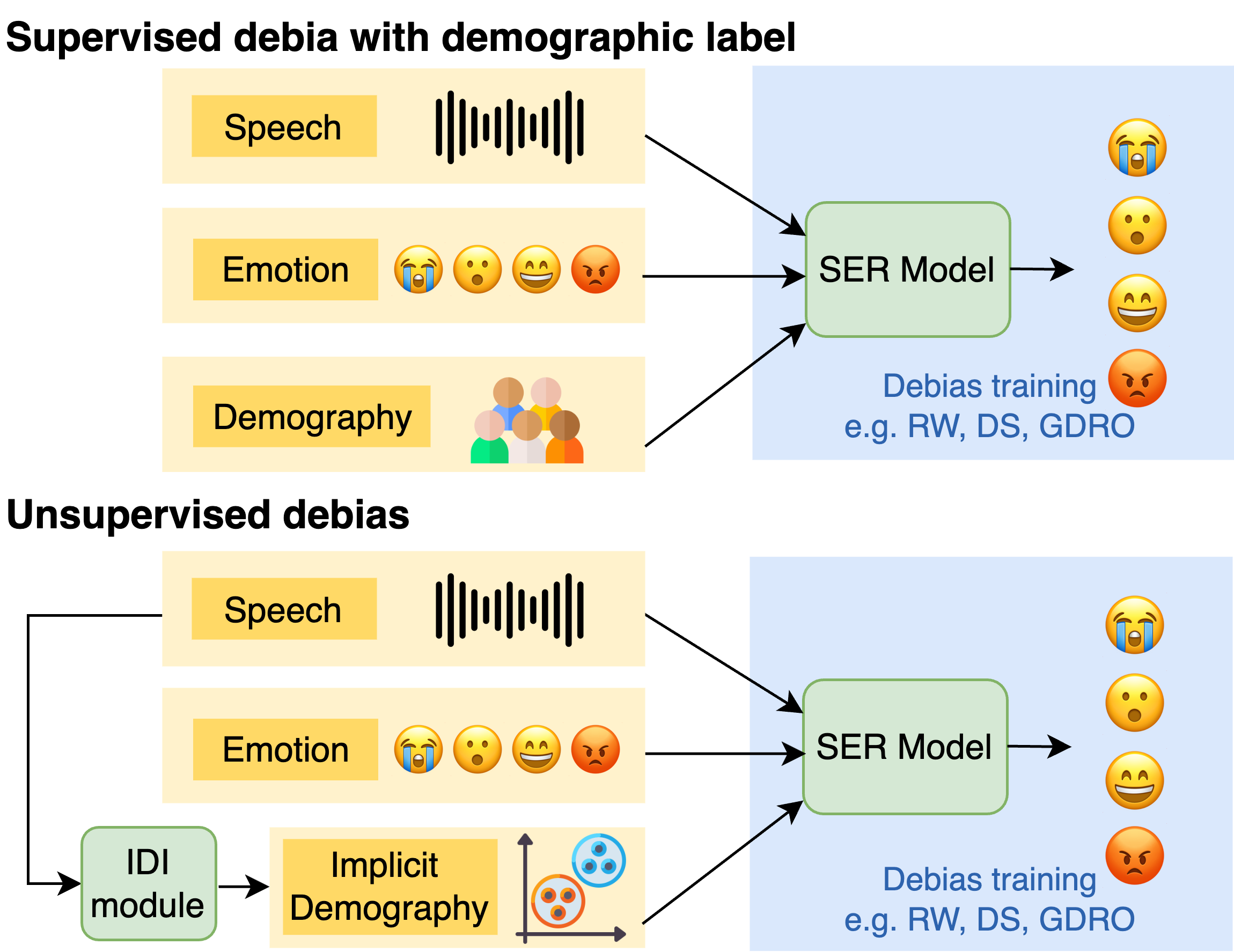} 
    \caption{Illustration of the proposed debias framework.}
    \label{fig:framework} 
\end{figure}

\section{Related Works}


\subsection{Debias by Supervised Learning}

Supervised learning approaches to mitigating gender bias in SER typically rely on explicitly annotated gender labels during training. Previous studies have implemented fairness constraints or modified training strategies to reduce performance disparities between gender subgroups. Chien et al. \cite{Chien_2023} employed fairness-constrained adversarial training to develop gender-fair representations, considering both speaker and rater gender to enhance the flexibility of emotion predictions. Similarly, Chien et al. \cite{Chien_2024} addressed biases originating from both speakers and raters by integrating fairness-aware learning techniques that combine adversarial training with multi-objective optimization, thereby producing gender-neutral emotion predictions. More recently, Upadhyay et al. \cite{Upadhyay_2025} demonstrated that while SER models may exhibit fairness within their source datasets, significant gender biases can emerge in cross-corpus scenarios. To mitigate this issue, they proposed the Combined Fairness Adaptation method, which leverages both adversarial and contrastive learning to align gender representations across domains, enhancing fairness without compromising recognition performance. However, these studies rely on gender labels, which are often difficult to obtain due to privacy concerns and data limitations. Additionally, they frequently simplify test sets by selecting or removing data to focus on a subset of emotions. Furthermore, these approaches typically treat SER as a single-label classification task, disregarding samples without consensus emotion labels in the test set.

In contrast to these prior studies, our approach retains all data and labels in the test set to provide a more accurate assessment of model performance. Moreover, we define SER as a multi-label task to better capture the complexity of emotion perception, as supported by recent SER research \cite{Chou_2024_v3,Wu_2024,Park_2024} and psychological insights \cite{Cowen_2021}. Emotion perception is inherently multifaceted, with speakers frequently conveying multiple emotions simultaneously. For example, an utterance may express both anger and sarcasm. Thus, multi-label classification presents a more suitable paradigm for SER than traditional single-label classification.

\subsection{Debias by Unsupervised Learning}

Chou et al. \cite{Chou_2025} present a method to mitigate individual speaker biases by employing unsupervised clustering on utterance embeddings from a pre-trained speaker verification model. This method effectively clusters utterances with similar characteristics, approximating true speaker identities without explicit identification. The study shows that incorporating these cluster identifiers enables the creation of a fairness-aware SER model that operates equitably across individual speakers. However, this work focuses on dimensional SER—such as arousal, valence, and dominance—and uses a small subset of the original emotion dataset, which, as Ferrer et al. \cite{Ferrer_2024} suggest, may lead to overestimated performance in testing. Therefore, we evaluate all experiments using the original test set.

\section{Methodology}
We propose a novel framework to mitigate subgroup disparities in gender in multi-label SER without relying on explicit demographic annotations. Central to our approach is the \emph{Implicit Demography Inference (IDI) Module}, which infers group labels from speech data by combining pseudo-labeling and unsupervised clustering. These inferred group labels are then utilized to guide a debias training process. An overview of the proposed framework is illustrated in Fig.~\ref{fig:framework}. In the following sections, we detail the architecture of our proposed IDI module and its integration with debias training techniques.

\subsection{Implicit Demography Inference (IDI) Module}
The IDI Module is designed to automatically infer latent group labels from speech features through one of two methods: pseudo-labeling and unsupervised clustering.

\textbf{Pseudo-labeling}: we leverage a pre-trained gender detection model \cite{Felix_2023}\footnote{\tiny https://huggingface.co/audeering/wav2vec2-large-robust-24-ft-age-gender} to generate pseudo gender labels for each input utterance. These labels serve as initial proxies, capturing potential demographic cues inherent in the speech signal. The pseudo-labels offer a straightforward means to incorporate demographic information without manual annotation. The model's reliability is assessed on the CREMA-D dataset (detailed in Sec. \ref{subsec:dataset}), achieving an accuracy of 94.4\%.

\textbf{Unsupervised Clustering}: We extract embeddings using the ECAPA-TDNN model \cite{Desplanques_2020} and apply K-Means clustering \cite{Lloyd_1982} (details about parameters are in Appendix Table~\ref{tab:appendix_sklearn}) to these embeddings. The cluster assignments derived from this process are treated as group labels, reflecting latent structures within the data that may correspond to demographic differences. We set the cluster sizes to 2, 4, 8, 16, and 32.

\subsection{Debias Training with Inferred Group Labels}

Once the IDI Module infers the group labels, they are incorporated into the debias training of the SER model. Specifically, the inferred labels partition the training data into subgroups, allowing us to impose fairness constraints that decrease performance disparities across these groups. We explore four debiasing methods that leverage these inferred group labels:

\textbf{Reweighting (RW)} \cite{kamiran2012data}: This approach adjusts the importance of each sample based on the frequency of its subgroup within each emotion class. The overall weighted loss is computed as follows:

\begin{equation}
\mathcal{L}_{RW} = \sum_{c}\sum_{g} \frac{n_c}{n_{c,g}} \, \mathcal{L}_{SER}^{(c,g)},
\end{equation}

where \(n_{c,g}\) is the number of samples in class \(c\) belonging to subgroup \(g\), reflects the inverse frequency of subgroup \(g\) in the specific emotion class. \(n_c\) is the number of samples in class \(c\). \(\mathcal{L}_{SER}^{(c, g)}\) denotes the SER loss in class \(c\) and subgroup \(g\).
    
\textbf{Downsampling (DS)} \cite{kamiran2012data}: This method balances subgroup distributions by equalizing the number of samples per group within each class. For each class, we determine the minimum number of utterances available among all groups and then randomly select that many samples from each group. This ensures that every subgroup is equally represented during training.
    
\textbf{Group Distribution Robust Optimization (GDRO)} \cite{gdro}: GDRO seeks to minimize the worst-case loss among all subgroups, ensuring robust performance across groups. The GDRO objective is formulated as:
\begin{equation}
    \mathcal{L}_{GDRO} = \max_{g} \, \mathcal{L}_{SER}^{(g)},
\end{equation}
which directly targets the subgroup with the highest loss.
    
\textbf{Group Aware DRO (GADRO)} \cite{sagawadistributionally}: Extending GDRO, GADRO incorporates an additional regularization term \(\frac{\lambda_{GD}}{\sqrt{n_g}}\) to penalize large discrepancies between subgroup losses, further promoting fairness. The loss function for GADRO is defined as:
\begin{equation}
    \mathcal{L}_{GADRO} = \max_{g} \, (\mathcal{L}_{SER}^{(g)} + \frac{\lambda_{GD}}{\sqrt{n_g}}),
\end{equation}
where \(\lambda_{GD}\) is a hyperparameter and \(n_g\) denotes the group size. In our experiment, we set \(\lambda_{GD}\) to 4.

\section{Experiments Setup}

We select the SSL-based SER model with WavLM base plus\footnote{\tiny https://huggingface.co/s3prl/converted\_ckpts/resolve/main/wavlm\_base\_plus.pt} \cite{chen2022wavlm} feature extractor and two linear layers as our primary backbone for the following experiments, because it ranks second on the SER leaderboard \cite{Wu_2024} while maintaining a lower parameter count compared to top-1 models, XLSR-1B. We use the class-balanced cross-entropy loss \cite{Cui_2019} as base SER loss $\mathcal{L}_{SER}$.

\subsection{Baseline}

We compare our proposed methods against several baseline models. \textbf{Random Baseline:} A naive model that assigns predictions uniformly at random, establishing a lower bound on SER performance. \textbf{Empirical Risk Minimization (ERM):} The standard SER model minimizes the overall  $\mathcal{L}_{SER}$ on the training data without any fairness constraints. ERM serves as our conventional reference model. \textbf{Existing Unsupervised Debiasing Methods:} We implement state-of-the-art debiasing approaches from computer vision that do not require explicit demographic labels, including LfF \cite{Nam_2020}, 
and DisEnt \cite{lee2021learning}. \textbf{PGDRO:} A method that leverages probabilistic pseudo-labels for demography inference \cite{Ghosal_2023}.


\subsection{Dataset}

\label{subsec:dataset}
We evaluate our method on the widely used CREMA-D database \cite{Cao_2014}, which contains 7,442 utterances (from 91 actors) in in English across six emotions (anger, disgust, fear, happy, neutral, sad). Each utterance is annotated by at least 7 raters and includes demographic metadata (gender, age, race, and ethnicity). For reproducibility, we use the audio-only annotations and adhere to the EMO-SUPERB splits \cite{Wu_2024}.

Most prior works on debiasing in SER have filtered or selectively sampled data for their experiments. Inspired by Ferrer et al. \cite{Ferrer_2024}, we simulate biased training data while evaluating models on the original test set to reflect real-world conditions in an unseen setting. To achieve this, we employ ad-hoc data pre-processing to manipulate the training data. First, we assign each sample to a single emotion category based on majority voting. Then, we follow the natural demographic trends in the CREMA-D dataset. In this study, we set the gender imbalance ratio to 1:20 (female to male or male to female). Appendix Table~\ref{tab:1_20_ratio_example} and Table~\ref{tab:other_data_distribution} provide a summary of this data distribution.







\begin{table}[t]
\centering
\fontsize{8}{9}\selectfont
\caption{Summary of results in SER accuracy (\textbf{F1} and \textbf{ACC}) and fairness  (\textbf{TPR$_{gap}$} and \textbf{DP$_{gap}$}) on gender. \textbf{Super.} specifies the level of supervision for debias. \textbf{\CIRCLE} indicates debias with ground truth gender labels. \textbf{\LEFTcircle} represents the pseudo-labeling approach, and \textbf{\Circle} refers to bias unsupervised methods. For unsupervised IDI, we use k-means with k=16. \textbf{Overall} averages the four IDI methods. The \textbf{Gain} represents the improvement compared to the \textbf{ERM} method. \textbf{$\uparrow$} indicates that higher values correspond to better performance; \textbf{$\downarrow$} indicates the opposite.}
\vspace{-2mm}
\begin{tabular}{@{\hspace{0.2cm}}c@{\hspace{0.2cm}}c@{\hspace{0.2cm}}c@{\hspace{0.2cm}}c@{\hspace{0.2cm}}c@{\hspace{0.2cm}}c@{\hspace{0.2cm}}}
\toprule
\textbf{Method}                     & \textbf{Super.}    & \textbf{F1$\uparrow$}    & \textbf{ACC$\uparrow$}   & \textbf{TPR$_{gap}$$\downarrow$} & \textbf{DP$_{gap}$$\downarrow$} \\ \midrule
ERM                        &          & 0.651 & 0.824 & 0.278   & 0.103  \\ 
Random                     &          & 0.360 & 0.509 & 0.060   & 0.015  \\ \midrule

LfF \cite{Nam_2020}        &   \Circle   & \textbf{0.655} & \underline{0.825} & 0.268   & 0.099  \\
DisEnt \cite{lee2021learning}       &  \Circle  & \underline{0.653} & \underline{0.825} & 0.261   & 0.094  \\ 
PGDRO \cite{Ghosal_2023}                     &  \textbf{\LEFTcircle}       & 0.643 & 0.817 & 0.318   & 0.121  \\ \specialrule{1pt}{1.4pt}{2.1pt}
DS \cite{kamiran2012data}       & \CIRCLE        & 0.598 & 0.804 & \textbf{0.094}   & \textbf{0.033}  \\ \arrayrulecolor{lightgray} \cmidrule(l){2-6} \arrayrulecolor{black} 
\multirow{2}{*}{DS-IDI} &  \textbf{\LEFTcircle}       & 0.580 & 0.811 & \underline{0.137}   & \underline{0.045}  \\  
                        & \Circle & 0.561 & 0.755 & 0.201   & 0.100  \\ \midrule
RW \cite{kamiran2012data}   & \CIRCLE        & 0.646 & 0.823 & 0.171   & 0.061  \\ \arrayrulecolor{lightgray} \cmidrule(l){2-6} \arrayrulecolor{black}
\multirow{2}{*}{RW-IDI} &  \textbf{\LEFTcircle}       & 0.647 & \textbf{0.826} & 0.197   & 0.071  \\
                           & \Circle & 0.642 & 0.817 & 0.237   & 0.091  \\ \midrule
GDRO \cite{gdro}      & \CIRCLE        & 0.627 & 0.796 & 0.199   & 0.086  \\  \arrayrulecolor{lightgray} \cmidrule(l){2-6} \arrayrulecolor{black}
\multirow{2}{*}{GDRO-IDI}  &  \textbf{\LEFTcircle}       & 0.632 & 0.807 & 0.229   & 0.088  \\
                           & \Circle & 0.639 & 0.803 & 0.228   & 0.104  \\ \midrule
GADRO \cite{sagawadistributionally}  & \CIRCLE        & 0.621 & 0.788 & 0.185   & 0.084  \\  \arrayrulecolor{lightgray} \cmidrule(l){2-6} \arrayrulecolor{black}
\multirow{2}{*}{GADRO-IDI}&  \textbf{\LEFTcircle}       & 0.631 & 0.807 & 0.229   & 0.088  \\ 
                           & \Circle & 0.638 & 0.804 & 0.223   & 0.098  \\  \specialrule{1pt}{1.4pt}{2.1pt}
\multirow{3}{*}{Overall} & \CIRCLE        & \textbf{0.623}   & 0.803   & \textbf{0.162}   & \textbf{0.066}   \\
                         & \LEFTcircle       & \textbf{0.623}   & \textbf{0.813}   & 0.198   & 0.073   \\
                         & \Circle & 0.620   & 0.795   & 0.222   & 0.098   \\ \midrule
\multirow{3}{*}{\textbf{Gain$\uparrow$}}    & \CIRCLE        & \textbf{-4.30\%} & -2.58\% & \textbf{41.64\%} & \textbf{35.92\%} \\
                         & \LEFTcircle       & -4.38\% & \textbf{-1.37\%} & 28.78\% & 29.13\% \\
                         & \Circle & -4.76\% & -3.55\% & 20.05\% & 4.61\% \\ \bottomrule
\end{tabular}
\label{tab:all_results}
\vspace{-5mm}
\end{table}


\subsection{Evaluation Metrics}

For assessing SER, we utilize two primary performance metrics: the macro-F1 score (\textbf{F1}) and Hamming accuracy (\textbf{ACC}). The macro-F1 score is computed by evaluating the F1 measure for each emotion category individually and then averaging the results, thereby handling class imbalances. Hamming accuracy, on the other hand, calculates the ratio of correctly predicted labels to the total number of labels, offering a comprehensive view of overall model performance. In accordance with the approach outlined in \cite{Wu_2024}, we binarize the prediction probabilities using a threshold of $1/|\mathcal{Y}|$, where $\mathcal{Y}$ denotes the set of all emotion classes.

In terms of fairness, our evaluation is based on two critical criteria: Equal Opportunity and Demographic Parity. The Equal Opportunity metric is determined by calculating the root mean square (RMS) of the gap in the true positive rate (TPR) across all emotion classes \cite{han-etal-2021-diverse, 10.1145/3287560.3287572}. This measure ensures that, given the actual positive instances, the likelihood of correctly predicting an emotion does not vary with protected attributes (e.g., gender, race). In addition, Demographic Parity requires that the probability of a positive prediction remains consistent across different demographic groups, independent of the true label.
We express this disparity mathematically as: \begin{equation} 
DP_{gap} = \sqrt{\sum_{\hat{y}\in\mathcal{Y}}\max_{g\in\mathcal{G}} DP(g, \hat{y})^2};
\end{equation}
\begin{equation} 
DP(g, \hat{y}) = \mathbb{E} \Big[P(\hat{y} = 1 \mid G = g) - P(\hat{y} = 1)\Big],
\end{equation} 
where $G$ represents a group, $\mathcal{G}$ represents the set of all groups, and $\hat{y}$ represents the emotion prediction. A higher value of DP$_{gap}$ implies a greater imbalance.

\section{Results and Analyses}
Table \ref{tab:all_results} presents the averaged results on performance and gender bias across five folds. We evaluate the effectiveness of our proposed methods and provide detailed analyses in the following sections.


\subsection{Performance and Effects of Pseudo-labeling}
Our results show that pseudo-labeling reduces subgroup disparities in multi-label SER. As seen in Table \ref{tab:all_results}, using pseudo-labels from a pre-trained gender detection model improves the TPR$_{gap}$ fairness metric by 28.78\% and DP by 29.13\% relative to the ERM baseline, with less than a 1.5\% drop in overall accuracy. Compared to PGDRO \cite{Ghosal_2023}, our modified reweighting strategy with pseudo-labels achieves a better balance between fairness and performance.

\subsection{Performance and Effects of Unsupervised Learning}


The number of clusters plays a critical role in the effectiveness of this method. Using too few clusters may result in overly coarse groupings that fail to capture the underlying nuances in the data, while employing too many clusters can fragment the data and introduce noise, leading to less meaningful subgroup distinctions. We found that a cluster size of 16 provides a balanced granularity, offering significant improvements in fairness metrics (20.05\% in TPR$_{gap}$ and 4.61\% in DP$_{gap}$) while maintaining a minor drop in overall performance (a 4.76\% decrease in F1 and a 3.55\% decrease in ACC), as shown in Table \ref{tab:all_results}. 

When compared to previously proposed unsupervised debiasing methods such as LfF and DisEnt, our approach exhibits a more favorable trade-off between fairness and SER performance. While these earlier methods yield minor performance improvements, they often incur few debias effects in TPR$_{gap}$ and DP$_{gap}$. In contrast, our method not only enhances fairness metrics significantly but also maintains higher recognition performance.

\begin{figure}[h]  
    \centering
    \includegraphics[width=0.5\textwidth]{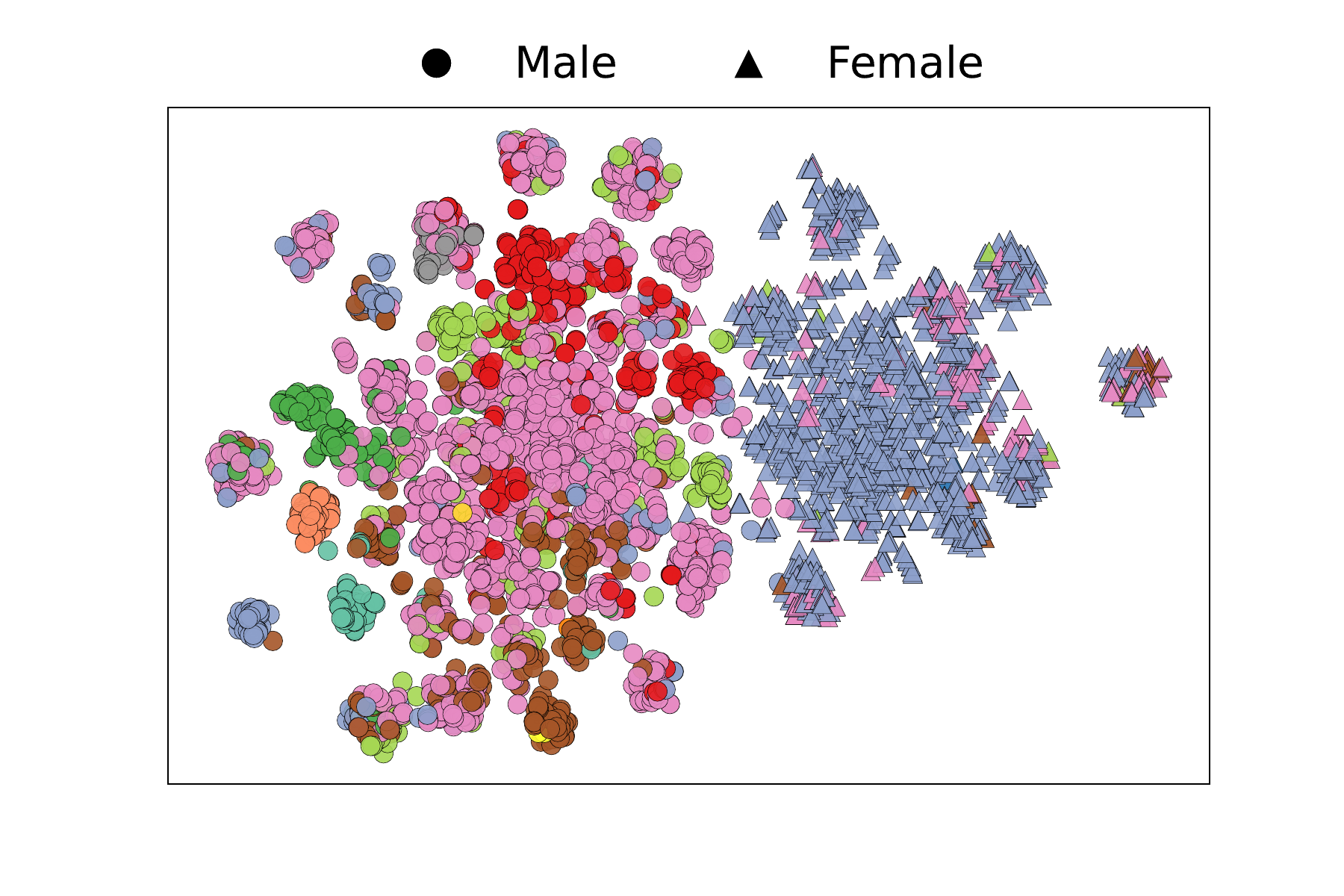} 
    \caption{Illustration of samples in Fold 5 using t-SNE (details about parameters are in Appendix Table~\ref{tab:appendix_sklearn}), with ECAPA-TDNN embeddings and a cluster size of 16. Circles and triangles represent male and female speakers, respectively, while different colors indicate different clusters.}
    \label{fig:tsne} 
\end{figure}

To further investigate the underlying reasons behind the method's effectiveness, we visualize the ECAPA-TDNN embeddings using t-SNE \cite{van2008visualizing}
, as shown in Fig.~\ref{fig:tsne}. 
The male and female samples are clearly separated, confirming that our clustering captures gender information and supports the scalability and label-free nature of our unsupervised approach.


\begin{table}[t]
\centering
\fontsize{8}{9}\selectfont
\caption{\small The results of unsupervised clustering on \textbf{Race} and \textbf{Age}. \textbf{G} denotes the number of subgroups within the demographic information. \textbf{K} refers to the number of clusters. \textbf{Gain} represents the performance improvement compared to the baseline model, \textbf{ERM}.}
\vspace{-3mm}
\begin{tabular}{@{\hspace{0.3cm}}c@{\hspace{0.3cm}}c|@{\hspace{0.3cm}}c@{\hspace{0.03cm}}c|@{\hspace{0.03cm}}c@{\hspace{0.03cm}}c@{\hspace{0.03cm}}c@{\hspace{0.03cm}}c@{\hspace{-0.03cm}}}
\toprule
\multicolumn{4}{c}{}           & \multicolumn{2}{c}{Race (G=3)} & \multicolumn{2}{c}{Age (G=3)} \\ \midrule
Model                  & K & F1$\uparrow$    & ACC$\uparrow$   & TPR$_{gap}$ & DP$_{gap}$ & TPR$_{gap}$ & DP$_{gap}$  \\ \toprule
ERM                    &         & 0.651 & 0.824 & 0.183             & 0.111           & 0.153            & 0.090           \\ \midrule
\multirow{2}{*}{DS}    & 16      & 0.561 & 0.755 & 0.142             & 0.119           & 0.134            & 0.095           \\
                       & 32      & 0.575 & 0.773 & 0.165             & 0.112           & 0.137            & 0.106           \\ \midrule
\multirow{2}{*}{RW}    & 16      & 0.642 & 0.817 & 0.147             & 0.100           & 0.140            & 0.079           \\
                       & 32      & 0.628 & 0.810 & 0.172             & 0.112           & 0.150            & 0.095           \\ \midrule
\multirow{2}{*}{GADRO} & 16      & 0.639 & 0.803 & 0.159             & 0.118           & 0.138            & 0.095           \\
                       & 32      & 0.640 & 0.806 & 0.162             & 0.109           & 0.137            & 0.096           \\ \midrule
\multirow{2}{*}{GDRO}  & 16      & 0.638 & 0.804 & 0.162             & 0.113           & 0.122            & 0.089           \\
                       & 32      & 0.629 & 0.800 & 0.172             & 0.111           & 0.126            & 0.090           \\ \midrule
\multirow{2}{*}{\textbf{Gain}$\uparrow$}  & 16      & -4.76\% & -3.55\% & \textbf{16.67\%}          & -1.35\%          & \textbf{12.75\%}          & 0.56\%          \\
                       & 32      & -5.07\% & -3.25\% & 8.33\%           & 0.00\%           & 10.13\%          & -7.50\%         \\ \bottomrule
\end{tabular}
\label{tab:others}
\vspace{-5mm}
\end{table}

\begin{figure}[h]
    \centering
    %
    \begin{subfigure}[b]{0.4\textwidth} 
        \centering
        \includegraphics[width=\textwidth]{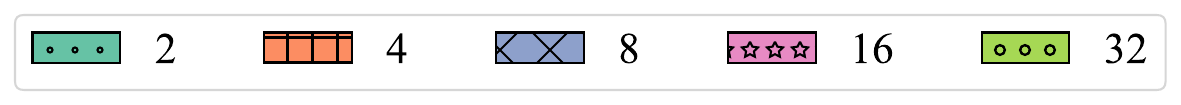} 
    \end{subfigure}
    \begin{subfigure}[b]{0.233\textwidth}
        \centering
        \includegraphics[width=\textwidth]{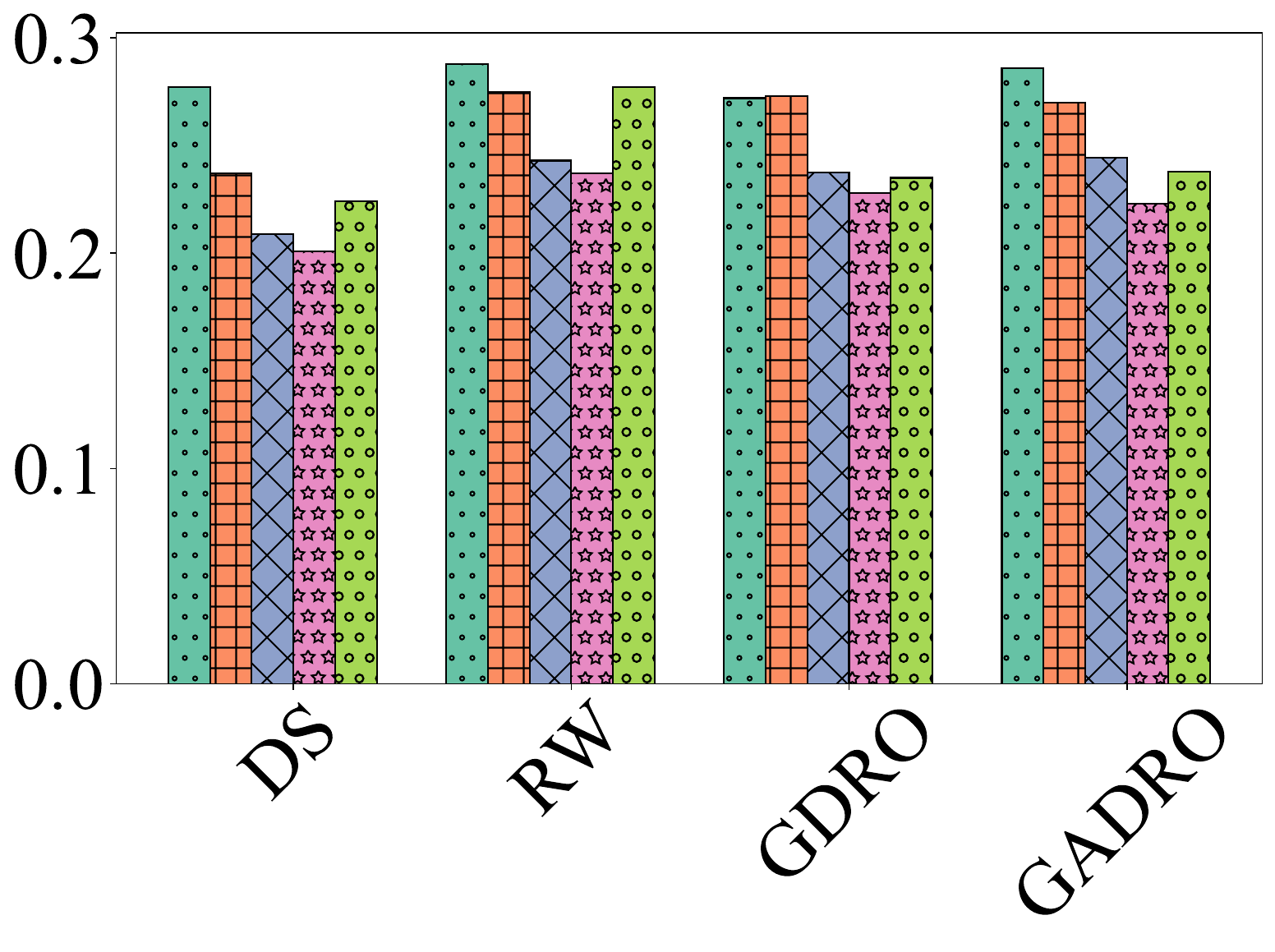}
        \label{fig:clusters_TPR}
    \end{subfigure}
    \hfill
    \begin{subfigure}[b]{0.233\textwidth}
        \centering
        \includegraphics[width=\textwidth]{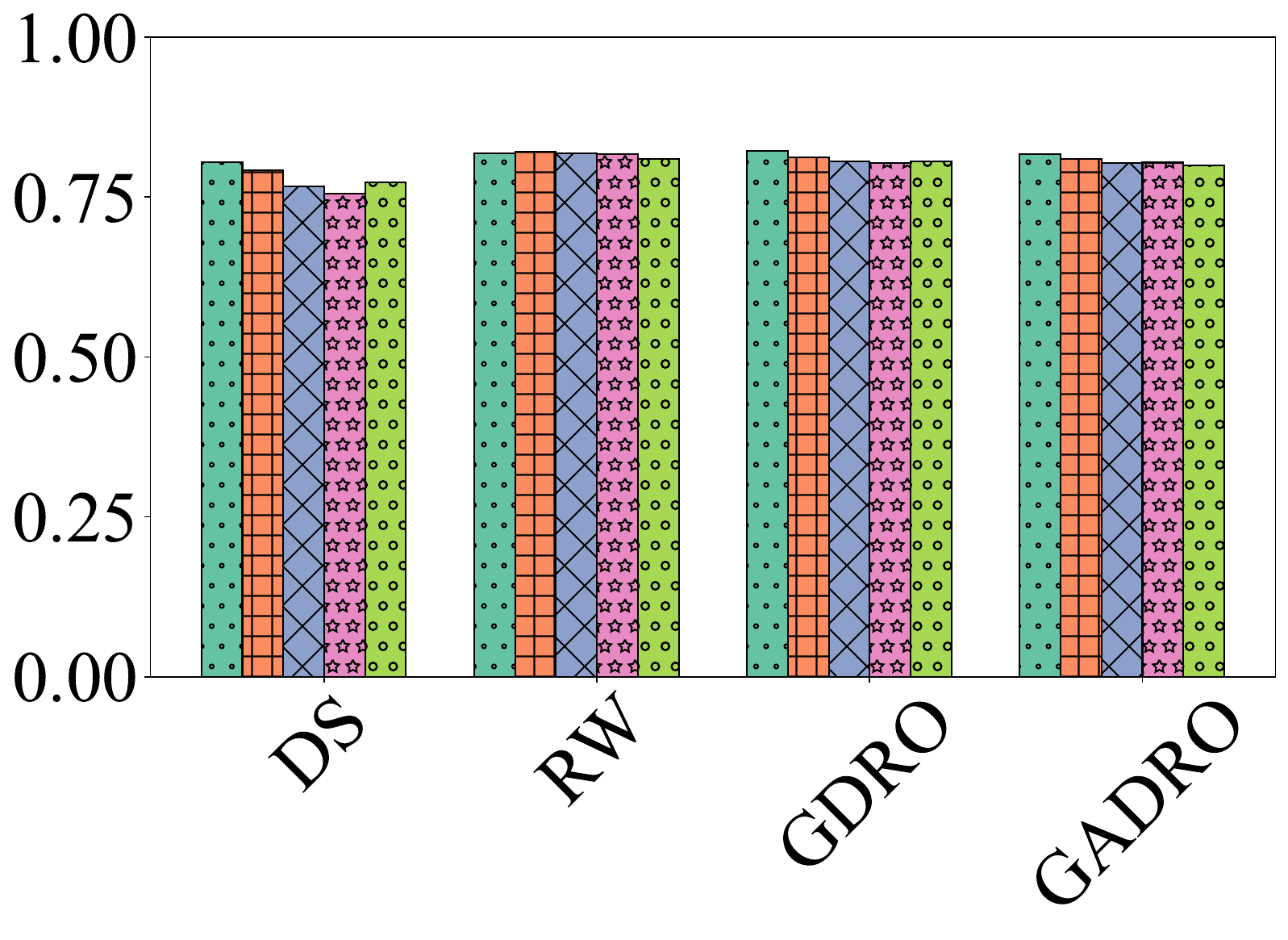}
        \label{fig:clusters_ACC}
    \end{subfigure}
    \caption{The left subfigure presents the  TPR$_{gap}$, while the right one depicts the Hamming accuracy. The bars illustrate the performance of the proposed unsupervised debiasing method, evaluated across four different techniques with varying cluster sizes.}
    \label{fig:clusters}
\end{figure}

\begin{figure}[h]  
    \centering
    
    \includegraphics[width=0.45\textwidth]{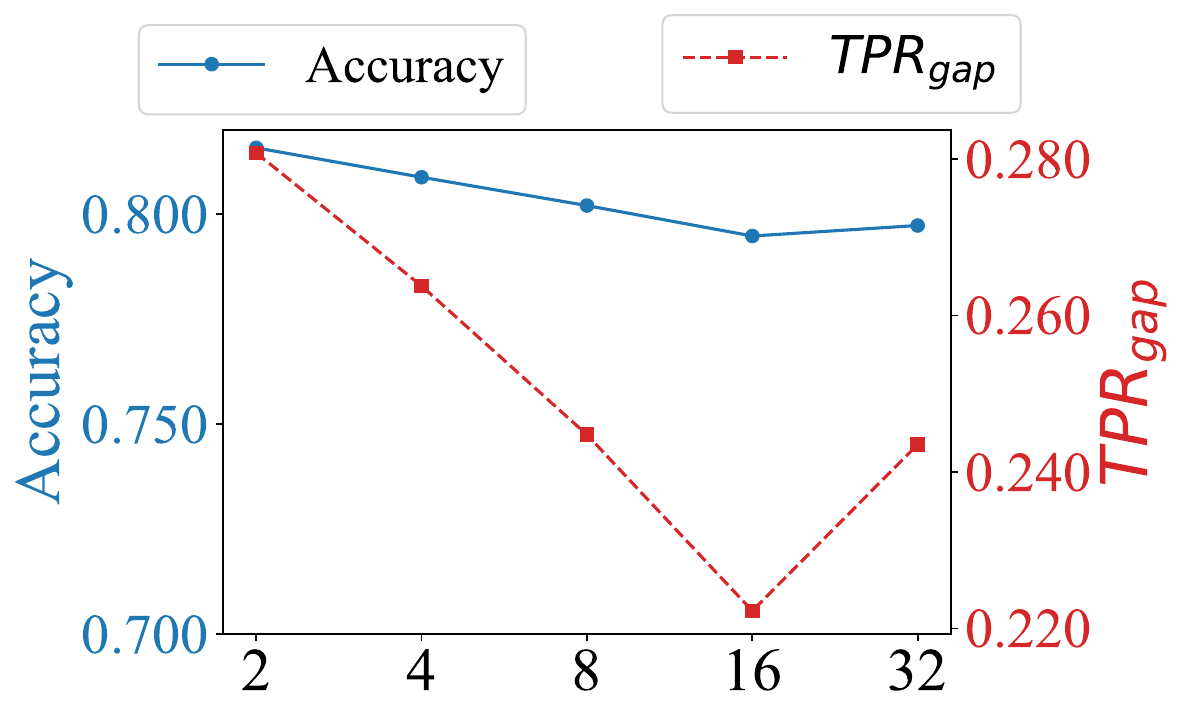} 
    \caption{Illustration of the averaged results across four models trained with the proposed unsupervised method, using different cluster sizes, in terms of Accuracy and TPR.}
    \label{fig:trends} 
\end{figure}

\subsection{Potential of Unsupervised Clustering}

We assess the applicability of unsupervised clustering to additional demographic attributes, as it does not rely on explicit demographic information. The CREMA-D dataset provides demographic metadata including speakers' age and race (specifically, Caucasian, African American, and Asian), and their distributions are detailed in Appendix Table \ref{tab:other_data_distribution}. In accordance with \cite{Goncalves_2025}, ages are classified into three subgroups:``Young" (20–35 years),``Middle" (36–49 years), and ``Elderly" (50–75 years).

Table \ref{tab:others} presents overall performance across subgroups, demonstrating that the unsupervised method with K=16 clusters improves TPR$_{gap}$ metric by over 12.75\%, while maintaining an SER performance decrease of 4.76\% and 3.55\% in F1 and ACC, respectively. This suggests that the method's performance is robust across different demographic attributes, highlighting its potential for broad applicability for subgroup disparities in multi-label SER tasks.

\section{Limitations}

While our approach demonstrates robust improvements in mitigating subgroup disparities, several limitations warrant discussion. First, our experiments are based on the CREMA-D dataset, which comprises acted emotional expressions; thus, the generalizability to naturalistic settings remains to be validated. Second, our methods depend heavily on the quality of pre-trained models—the gender detection model for pseudo-labeling and the ECAPA-TDNN for embedding extraction. More pre-trained models can be explored.

\section{Conclusion and Future Work}

This work empirically examines various debiasing methods in supervised, pseudo-label, and unsupervised learning to mitigate subgroup disparities without explicit demographic information. We leverage pseudo-labels generated by a pre-trained model and introduce an unsupervised clustering approach that integrates with existing debiasing techniques originally designed for demographic labels. Our findings demonstrate that pseudo-label learning improves TPR and DP metrics by 28.78\% and 29.13\%, respectively, with only a minor 4.38\% degradation in SER accuracy. Similarly, our proposed unsupervised method achieves 20.05\% and 4.61\% improvements in TPR and DP metrics, respectively, with just a 4.76\% drop in SER performance. For future work, we will explore more possible bias attributes like content \cite{10832317}, language \cite{lin2025improvingspeechemotionrecognition}, or speaking style \cite{9747897}.

\section{Acknowledgement}
We thank Taiwan’s National Center for High-Performance Computing (NCHC) at the National Applied Research Laboratories (NARLabs) for providing the computational and storage resources.

\bibliographystyle{IEEEtran}
\bibliography{mybib}

\begin{thebibliography}{10}
\providecommand{\url}[1]{#1}
\csname url@samestyle\endcsname
\providecommand{\newblock}{\relax}
\providecommand{\bibinfo}[2]{#2}
\providecommand{\BIBentrySTDinterwordspacing}{\spaceskip=0pt\relax}
\providecommand{\BIBentryALTinterwordstretchfactor}{4}
\providecommand{\BIBentryALTinterwordspacing}{\spaceskip=\fontdimen2\font plus
\BIBentryALTinterwordstretchfactor\fontdimen3\font minus \fontdimen4\font\relax}
\providecommand{\BIBforeignlanguage}[2]{{%
\expandafter\ifx\csname l@#1\endcsname\relax
\typeout{** WARNING: IEEEtran.bst: No hyphenation pattern has been}%
\typeout{** loaded for the language `#1'. Using the pattern for}%
\typeout{** the default language instead.}%
\else
\language=\csname l@#1\endcsname
\fi
#2}}
\providecommand{\BIBdecl}{\relax}
\BIBdecl

\bibitem{Thiripurasundari_2024}
D.~Thiripurasundari, K.~Bhangale, V.~Aashritha, S.~Mondreti, and M.~Kothandaraman, ``{Speech emotion recognition for human–computer interaction},'' \emph{Int. J. Speech Technol.}, vol.~27, no.~3, p. 817–830, Aug. 2024.

\bibitem{Daniel_2022}
D.~Varona and J.~L. Suárez, ``{Discrimination, Bias, Fairness, and Trustworthy AI},'' \emph{Applied Sciences}, vol.~12, no.~12, 2022.

\bibitem{Carolyn_2023}
C.~Ashurst and A.~Weller, ``{Fairness Without Demographic Data: A Survey of Approaches},'' in \emph{Proceedings of the 3rd ACM Conference on Equity and Access in Algorithms, Mechanisms, and Optimization}, ser. EAAMO '23.\hskip 1em plus 0.5em minus 0.4em\relax New York, NY, USA: Association for Computing Machinery, 2023.

\bibitem{Lin_2024}
Y.-C. Lin, H.~Wu, H.-C. Chou, C.-C. Lee, and H.~yi~Lee, ``{Emo-bias: A Large Scale Evaluation of Social Bias on Speech Emotion Recognition},'' in \emph{Interspeech 2024}, 2024, pp. 4633--4637.

\bibitem{Gorrostieta_2019}
C.~Gorrostieta, R.~Lotfian, K.~Taylor, R.~Brutti, and J.~Kane, ``{Gender De-Biasing in Speech Emotion Recognition},'' in \emph{Interspeech 2019}, 2019, pp. 2823--2827.

\bibitem{lin24b_interspeech}
Y.-C. Lin, T.-Q. Lin, H.-C. Lin, A.~T. Liu, and H.~yi~Lee, ``{On the social bias of speech self-supervised models},'' in \emph{Interspeech 2024}, 2024, pp. 4638--4642.

\bibitem{Chien_2023}
W.-S. Chien and C.-C. Lee, ``{Achieving Fair Speech Emotion Recognition via Perceptual Fairness},'' in \emph{ICASSP 2023 - 2023 IEEE International Conference on Acoustics, Speech and Signal Processing (ICASSP)}, 2023, pp. 1--5.

\bibitem{Chien_2024}
W.-S. Chien, S.~G. Upadhyay, and C.-C. Lee, ``{Balancing Speaker-Rater Fairness for Gender-Neutral Speech Emotion Recognition},'' in \emph{ICASSP 2024 - 2024 IEEE International Conference on Acoustics, Speech and Signal Processing (ICASSP)}, 2024, pp. 11\,861--11\,865.

\bibitem{10446326}
A.~Koudounas, E.~Pastor, G.~Attanasio, L.~de~Alfaro, and E.~Baralis, ``Prioritizing data acquisition for end-to-end speech model improvement,'' in \emph{ICASSP 2024 - 2024 IEEE International Conference on Acoustics, Speech and Signal Processing (ICASSP)}, 2024, pp. 7000--7004.

\bibitem{zhang22n_interspeech}
Y.~Zhang, Y.~Zhang, B.~Halpern, T.~Patel, and O.~Scharenborg, ``Mitigating bias against non-native accents,'' in \emph{Interspeech 2022}, 2022, pp. 3168--3172.

\bibitem{Chou_2024_v3}
H.-C. Chou, H.~Wu, L.~Goncalves, S.-G. Leem, A.~Salman, C.~Busso, H.-Y. Lee, and C.-C. Lee, ``{Embracing Ambiguity And Subjectivity Using The All-Inclusive Aggregation Rule For Evaluating Multi-Label Speech Emotion Recognition Systems},'' in \emph{2024 IEEE Spoken Language Technology Workshop (SLT)}, 2024, pp. 502--509.

\bibitem{Wu_2024}
H.~Wu \emph{et~al.}, ``{Open-Emotion: A Reproducible EMO-Superb For Speech Emotion Recognition Systems},'' in \emph{2024 IEEE Spoken Language Technology Workshop (SLT)}, 2024, pp. 510--517.

\bibitem{Cowen_2021}
A.~S. Cowen and D.~Keltner, ``{Semantic Space Theory: A Computational Approach to Emotion},'' \emph{Trends in Cognitive Sciences}, vol.~25, no.~2, pp. 124--136, 2021.

\bibitem{Desplanques_2020}
B.~Desplanques, J.~Thienpondt, and K.~Demuynck, ``{ECAPA-TDNN: Emphasized Channel Attention, Propagation and Aggregation in TDNN Based Speaker Verification},'' in \emph{Interspeech 2020}, 2020, pp. 3830--3834.

\bibitem{Cao_2014}
H.~Cao, D.~G. Cooper, M.~K. Keutmann, R.~C. Gur, A.~Nenkova, and R.~Verma, ``{CREMA-D: Crowd-Sourced Emotional Multimodal Actors Dataset},'' \emph{IEEE Transactions on Affective Computing}, vol.~5, no.~4, pp. 377--390, 2014.

\bibitem{Upadhyay_2025}
S.~G. Upadhyay, W.-S. Chien, and C.-C. Lee, ``{Is It Still Fair? Investigating Gender Fairness in Cross-Corpus Speech Emotion Recognition},'' in \emph{ICASSP 2025 - 2025 IEEE International Conference on Acoustics, Speech and Signal Processing (ICASSP)}, 2025.

\bibitem{Park_2024}
S.~Park, B.~Jeon, S.~Lee, and J.~Yoon, ``{Multi-Label Emotion Recognition of Korean Speech Data Using Deep Fusion Models},'' \emph{Applied Sciences}, vol.~14, no.~17, 2024.

\bibitem{Chou_2025}
H.-C. Chou, H.~Wu, and C.-C. Lee, ``{Stimulus Modality Matters: Impact of Perceptual Evaluations from Different Modalities on Speech Emotion Recognition System Performance},'' in \emph{ICASSP 2025 - 2025 IEEE International Conference on Acoustics, Speech and Signal Processing (ICASSP)}, 2025.

\bibitem{Ferrer_2024}
L.~Ferrer, O.~Scharenborg, and T.~Bäckström, ``{Good practices for evaluation of machine learning systems},'' 2024.

\bibitem{Felix_2023}
F.~Burkhardt, J.~Wagner, H.~Wierstorf, F.~Eyben, and B.~Schuller, ``{Speech-based Age and Gender Prediction with Transformers},'' in \emph{Speech Communication; 15th ITG Conference}, 2023, pp. 46--50.

\bibitem{Lloyd_1982}
S.~Lloyd, ``{Least squares quantization in PCM},'' \emph{IEEE Transactions on Information Theory}, vol.~28, no.~2, pp. 129--137, 1982.

\bibitem{kamiran2012data}
F.~Kamiran and T.~Calders, ``\BIBforeignlanguage{English}{{Data preprocessing techniques for classification without discrimination}},'' \emph{\BIBforeignlanguage{English}{Knowledge and Information Systems}}, vol.~33, no.~1, pp. 1--33, 2012.

\bibitem{gdro}
A.~Ben-Tal, D.~den Hertog, A.~De~Waegenaere, B.~Melenberg, and G.~Rennen, ``{Robust Solutions of Optimization Problems Affected by Uncertain Probabilities},'' \emph{Management Science}, vol.~59, no.~2, pp. 341--357, 2013.

\bibitem{sagawadistributionally}
S.~Sagawa*, P.~W. Koh*, T.~B. Hashimoto, and P.~Liang, ``{Distributionally Robust Neural Networks for Group Shifts: On the Importance of Regularization for Worst-Case Generalization.}'' in \emph{International Conference on Learning Representations}, 2020.

\bibitem{chen2022wavlm}
S.~Chen \emph{et~al.}, ``{WavLM: Large-Scale Self-Supervised Pre-Training for Full Stack Speech Processing},'' \emph{IEEE Journal of Selected Topics in Signal Processing}, 2022.

\bibitem{Cui_2019}
Y.~Cui, M.~Jia, T.-Y. Lin, Y.~Song, and S.~Belongie, ``{Class-Balanced Loss Based on Effective Number of Samples},'' in \emph{Proceedings of the IEEE/CVF Conference on Computer Vision and Pattern Recognition (CVPR)}, 2019.

\bibitem{Nam_2020}
J.~Nam, H.~Cha, S.~Ahn, J.~Lee, and J.~Shin, ``Learning from failure: training debiased classifier from biased classifier,'' in \emph{Proceedings of the 34th International Conference on Neural Information Processing Systems}, Red Hook, NY, USA, 2020.

\bibitem{lee2021learning}
J.~Lee, E.~Kim, J.~Lee, J.~Lee, and J.~Choo, ``{Learning Debiased Representation via Disentangled Feature Augmentation},'' in \emph{Advances in Neural Information Processing Systems}, 2021.

\bibitem{Ghosal_2023}
S.~S. Ghosal and Y.~Li, ``{Distributionally Robust Optimization with Probabilistic Group},'' \emph{Proceedings of the AAAI Conference on Artificial Intelligence}, vol.~37, no.~10, pp. 11\,809--11\,817, Jun. 2023.

\bibitem{han-etal-2021-diverse}
X.~Han, T.~Baldwin, and T.~Cohn, ``Diverse adversaries for mitigating bias in training,'' in \emph{Proceedings of the 16th Conference of the European Chapter of the Association for Computational Linguistics: Main Volume}, 2021.

\bibitem{10.1145/3287560.3287572}
M.~De-Arteaga, A.~Romanov, H.~Wallach, J.~Chayes, C.~Borgs, A.~Chouldechova, S.~Geyik, K.~Kenthapadi, and A.~T. Kalai, ``Bias in bios: A case study of semantic representation bias in a high-stakes setting,'' in \emph{Proceedings of the Conference on Fairness, Accountability, and Transparency}.\hskip 1em plus 0.5em minus 0.4em\relax Association for Computing Machinery, 2019.

\bibitem{van2008visualizing}
L.~van~der Maaten and G.~Hinton, ``{Visualizing Data using t-SNE},'' \emph{Journal of Machine Learning Research}, vol.~9, no.~86, pp. 2579--2605, 2008.

\bibitem{Goncalves_2025}
L.~Goncalves, H.-C. Chou, A.~N. Salman, C.-C. Lee, and C.~Busso, ``{Jointly Learning From Unimodal and Multimodal-Rated Labels in Audio-Visual Emotion Recognition},'' \emph{IEEE Open Journal of Signal Processing}, vol.~6, pp. 165--174, 2025.

\bibitem{10832317}
Y.-C. Lin \emph{et~al.}, ``Listen and speak fairly: a study on semantic gender bias in speech integrated large language models,'' in \emph{2024 IEEE Spoken Language Technology Workshop (SLT)}, 2024.

\bibitem{lin2025improvingspeechemotionrecognition}
\BIBentryALTinterwordspacing
H.-C. Lin, Y.-C. Lin, H.-C. Chou, and H.~yi~Lee, ``Improving speech emotion recognition in under-resourced languages via speech-to-speech translation with bootstrapping data selection,'' 2025. [Online]. Available: \url{https://arxiv.org/abs/2409.10985}
\BIBentrySTDinterwordspacing

\bibitem{9747897}
Y.~Meng, Y.-H. Chou, A.~T. Liu, and H.-y. Lee, ``Don't speak too fast: The impact of data bias on self-supervised speech models,'' in \emph{ICASSP 2022 - 2022 IEEE International Conference on Acoustics, Speech and Signal Processing (ICASSP)}, 2022.

\end{thebibliography}

\clearpage
\small
\appendix
\begin{center}
    \Large\bfseries Supplementary Material 
\end{center}

\renewcommand{\thetable}{A\arabic{table}} 
\renewcommand{\thefigure}{A\arabic{figure}} 
\setcounter{table}{0}  
\setcounter{figure}{0} 

The Supplementary Material furnishes a comprehensive account of our experimental setup. Table \ref{tab:kmeans_params} and \ref{tab:tsne_params} specify the implementations used for K-Means clustering and t-SNE visualization, respectively. Additionally, Table~\ref{tab:1_20_ratio_example} presents the data distribution at a 1:20 ratio for Fold 1 of the CREMA-D emotion database, and Table~\ref{tab:other_data_distribution} details the distribution of emotions, gender, race, and age across all five folds of the database.

\section{Implementation Details}
We use the Adam optimizer with a 0.0001 learning rate. The batch size and epoch are set as 32 and 50, respectively. 
We choose the best models according to the lowest value of the class-balanced cross-entropy loss on the development set. We use two Nvidia Tesla V100 GPUs with 32 GB of memory. The total number of GPU hours is around 500.

\begin{table}[h]
\centering
\fontsize{8}{10}\selectfont
\caption{\small Detailed configuration of the K-Means clustering algorithm as implemented in scikit-learn v1.2.2. This table lists every parameter we controlled.}
\label{tab:kmeans_params}
\begin{tabular}{@{}l l@{}}
\toprule
\textbf{Parameter}           & \textbf{Value}               \\ 
\midrule
Initialization                & \texttt{k-means++ (greedy)}  \\
Number of clusters           & \texttt{[2, 4, 8, 16, 32]}    \\
Random seed                  & 42                            \\
Maximum iterations           & 1 000                         \\
Mini-batch size              & 32                            \\
Compute final labels         & \texttt{True}                 \\
Reassignment ratio           & 0.01                          \\
Convergence tolerance        & $1 \times 10^{-4}$                      \\
Algorithm                     & Lloyd’s method                \\
\bottomrule
\end{tabular}
\end{table}

\begin{table}[h]
\centering
\fontsize{8}{10}\selectfont
\caption{\small Detailed configuration of the t-SNE embedding parameters used with scikit-learn v1.2.2.}
\label{tab:tsne_params}
\begin{tabular}{@{}l l@{}}
\toprule
\textbf{Parameter}                  & \textbf{Value}                   \\ 
\midrule
Initialization                       & PCA                              \\
Neighbor-search method               & Barnes–Hut                       \\
Random seed                          & 42                               \\
Target dimensionality                & 2                                \\
Perplexity                           & 40                               \\
Distance metric                      & Euclidean                        \\
Iterations without progress before abort & 300                          \\
Learning rate                        & \texttt{auto}                    \\
\bottomrule
\end{tabular}
\end{table}

\begin{table}[h]
\centering
\fontsize{8}{9}\selectfont
\caption{\small The table presents the manipulated biased data distribution at a 1:20 ratio in Fold 1.}
\vspace{-3mm}
\begin{tabular}{@{}ccccc@{}}
\toprule
Emotion                  & Gender & Development & Test & Train \\ \midrule
\multirow{2}{*}{Angry}   & Female & 4           & 102  & 12    \\
                         & Male   & 93          & 58   & 240   \\ \midrule
\multirow{2}{*}{Disgust} & Female & 1           & 41   & 4     \\
                         & Male   & 34          & 18   & 86    \\ \midrule
\multirow{2}{*}{Fear}    & Female & 27          & 55   & 127   \\
                         & Male   & 1           & 20   & 6     \\ \midrule
\multirow{2}{*}{Happy}   & Female & 24          & 35   & 102   \\
                         & Male   & 1           & 15   & 5     \\ \midrule
\multirow{2}{*}{Neutral} & Female & 19          & 263  & 54    \\
                         & Male   & 384         & 250  & 1080  \\ \midrule
\multirow{2}{*}{Sad}     & Female & 41          & 35   & 58    \\
                         & Male   & 2           & 7    & 2     \\ \bottomrule
\end{tabular}
\label{tab:1_20_ratio_example}
\vspace{-3mm}
\end{table}

\begin{table*}[b]
\centering
\fontsize{8}{9}\selectfont
\caption{\small Table summarizes detailed data distribution at 1:20 across emotions, gender, and age. \textbf{AA.} means African American. Note that an utterance may be associated with multiple emotions based on the distributional labels, using the threshold, $1 / 6$. \textbf{Dev.} means Development.}
\begin{tabular}{@{\hspace{0.1cm}}c@{\hspace{0.1cm}}|c|@{\hspace{0.1cm}}r@{\hspace{0.1cm}}r@{\hspace{0.1cm}}r@{\hspace{0.1cm}}r@{\hspace{0.1cm}}r@{\hspace{0.1cm}}r@{\hspace{0.1cm}}|r@{\hspace{0.1cm}}r@{\hspace{0.1cm}}|r@{\hspace{0.1cm}}r@{\hspace{0.1cm}}r@{\hspace{0.1cm}}|r@{\hspace{0.1cm}}r@{\hspace{0.1cm}}r@{\hspace{0.1cm}}}
\toprule
\multirow{2}{*}{Fold} & \multirow{2}{*}{Split Sets} & \multicolumn{6}{c|}{Emotion}                    & \multicolumn{2}{c|}{Gender} & \multicolumn{3}{c}{Race}             & \multicolumn{3}{c}{Age} \\ \cmidrule(l){3-16} 
                      &                             & Angry & Sad & Disgust & Fear & Neutral & Happy & Male        & Female       & Caucasian & AA. & Asian & Young  & Middle  & Elderly  \\ \toprule
\multirow{3}{*}{1}    & Train                       & 252   & 60  & 90      & 133  & 1134    & 107   & 1419        & 357          & 1070      & 466              & 229   & 927    & 449     & 400  \\ 
                      & Dev.                  & 97    & 43  & 35      & 28   & 403     & 25    & 515         & 116          & 476       & 109              & 46    & 365    & 151     & 115  \\
                      & Test                        & 160   & 42  & 59      & 75   & 513     & 50    & 574         & 889          & 1217      & 246              & 0     & 895    & 322     & 246  \\ \midrule
\multirow{3}{*}{2}    & Train                       & 227   & 77  & 77      & 135  & 1025    & 111   & 1282        & 370          & 1089      & 398              & 165   & 1028   & 266     & 358  \\
                      & Dev.                 & 85    & 19  & 31      & 55   & 370     & 32    & 467         & 125          & 349       & 170              & 62    & 188    & 268     & 136  \\
                      & Test                        & 147   & 55  & 55      & 69   & 590     & 38    & 896         & 574          & 1066      & 328              & 76    & 732    & 410     & 328  \\ \midrule
\multirow{3}{*}{3}    & Train                       & 225   & 105 & 85      & 124  & 1013    & 113   & 1276        & 389          & 1088      & 377              & 200   & 1059   & 296     & 310  \\
                      & Dev.                  & 99    & 15  & 28      & 38   & 415     & 23    & 520         & 98           & 470       & 135              & 13    & 334    & 121     & 163  \\
                      & Test                        & 143   & 36  & 43      & 96   & 605     & 48    & 820         & 656          & 984       & 328              & 82    & 656    & 574     & 246  \\ \midrule
\multirow{3}{*}{4}    & Train                       & 243   & 99  & 86      & 141  & 1036    & 94    & 1315        & 384          & 1187      & 390              & 107   & 839    & 511     & 349  \\
                      & Dev.                  & 67    & 25  & 30      & 38   & 347     & 51    & 428         & 130          & 245       & 159              & 154   & 404    & 56      & 98   \\
                      & Test                        & 154   & 30  & 48      & 76   & 609     & 37    & 900         & 574          & 900       & 492              & 82    & 738    & 246     & 490  \\ \midrule
\multirow{3}{*}{5}    & Train                       & 282   & 78  & 95      & 122  & 1189    & 80    & 1504        & 342          & 1290      & 425              & 117   & 888    & 543     & 415  \\
                      & Dev.                  & 60    & 36  & 18      & 57   & 262     & 36    & 330         & 139          & 364       & 105              & 0     & 277    & 88      & 104  \\
                      & Test                        & 136   & 30  & 48      & 59   & 663     & 60    & 738         & 817          & 817       & 410              & 328   & 1145   & 246     & 164  \\ \bottomrule 
\end{tabular}
\label{tab:other_data_distribution}
\end{table*}

\end{document}